# THE DESIGN OF A STREAMING ANALYTICAL WORKFLOW FOR PROCESSING MASSIVE TRANSIT FEEDS

Hung Cao and Monica Wachowicz

People in Motion Lab
University of New Brunswick
15 Dineen Drive, Fredericton, NB. E3B 5A3, Canada
(hcao3, monicaw)@unb.ca

**ABSTRACT:**

Retrieving and analyzing transit feeds relies on working with analytical workflows that can handle the massive volume of data streams that are relevant to understand the dynamics of transit networks which are entirely deterministic in the geographical space in which they takes place. In this paper, we consider the fundamental issues in developing a streaming analytical workflow for analyzing the continuous arrival of multiple, unbounded transit data feeds for automatically processing and enriching them with additional information containing higher level concepts accordingly to a particular mobility context. This workflow consists of three tasks: (1) stream data retrieval for creating time windows; (2) data cleaning for handling missing data, overlap data or redundant data; and (3) data contextualization for computing actual arrival and departure times as well as the stops and moves during a bus trip, and also performing mobility context computation. The workflow was implemented in a Hadoop cloud ecosystem using data streams from the CODIAC Transit System of the city of Moncton, NB. The *Map()* function of MapReduce is used to retrieve and bundle data streams into numerous clusters which are subsequently handled in a parallel manner by the *Reduce()* function in order to execute the data contextualization step. The results validate the need for cloud computing for achieving high performance and scalability, however, due to the delay in computing and networking, it is clear that data cleaning tasks should not only be deployed using a cloud environment, paving the way to combine it with fog computing in the near future.

## 1. INTRODUCTION

Streaming analytical workflows are designed to process data streams that are an unbounded sequence of tuples that usually have a high data input rate and are open to ad-hoc continuous queries. These analytical workflows apply complex processing tasks on most likely out-of-order tuples coming from a variety of different sources. Moreover, a streaming analytical workflow is needed to support the continuous computation of tuples flowing through different processing tasks. These tasks have been previously developed for network monitoring (Gupta et al., 2016), fraud detection (Rajeshwari U and Babu, 2016), data warehouse augmentation (Meehan and Zdonik, 2017) and risk management (Puthal et al., 2017). Although many transit authorities have installed GPS receivers and networking capabilities for transmitting stream data in real-time from their transit vehicles, previous research on streaming analytical workflows using transit feeds was not found, despite of our thorough review of the literature.

From a conceptual perspective, streaming analytical workflows play an important role in exploring data streams by searching for a hypothesis and proposing mobility contexts. Time is an important dimension of a streaming analytical workflow, and different models have been proposed in the literature to perform processing tasks in unbounded data streams, including landmark windows (Leung et al., 2013), sliding windows (Lee et al., 2014), and tilted windows (Giannella et al., 2003). In contrast, the space dimension has been neglected so far, even though data streams are being generated over large transit networks with high spatial granularity.

From an implementation perspective, a streaming analytical workflow requires (1) a pre-build connector that supports data connectivity to communicate with several data stream sources; (2) a low-latency database for storing live data streams in-memory, (3) a streaming processing environment for supporting aggregation, contextualization, and transformation techniques that are needed for supporting mobility analytics. The research challenge is to design a platform that can (1) perform analytical tasks without human intervention; and (2) cope with the processing of massive unbounded data streams where the data flow rate of the input tuples may overthrow the query processor.

This research paper aims to design a streaming analytical workflow taking into account both temporal and spatial semantics of data streams in order to support the processing of continuous computation of tuples flowing through three analytical tasks which are characterised as data ingestion, data cleaning and data contextualization. Each analytical task consists of several automated steps that are designed for processing massive transit feeds without human intervention.

This paper is organized as follows. Section 2 describes previous research work on data analysis of transit feeds. Section 3 introduces our streaming analytical workflow and in Section 4 the Hadoop cloud-based processing architecture is explained. The implementation using data streams from the CODIAC Transit System of the city of Moncton, NB is described in Section 5. Finally, the conclusions and future research recommendations are given in Section 6.

## 2. RELATED WORK

Analytics performed over data streams could potentially revolutionize transit network services that will be able to adapt at near real time to current or expected contexts, implementing real-time operation controls and recommender systems. (Huang et al.,

2014) provides the statistics of the Beijing Transportation Department that is already reaching a data ingestion of 15,000 GPS records per second for 30,000 buses. As we are moving towards the Internet of Moving Things, streaming analytics tasks will play a critical role in how data streams need to be processed, making manually batch processing based analytics to become obsolete when analyzing transit feeds of large transit networks.

Most of the research work found in analyzing transit feeds is based on manually batch processing using a cloud platform. (Poonawala et al., 2016) proposes the City in Motion (CiM) system built on top of a Hadoop-based platform for performing data ingestion using Apache Flume and Kafka. The data contains mobile network information including the location of a call, the mobile data usage, and network events (e.g. network congestion and call drop). Although hundreds of millions to over a billion tuples per day are stored in the CiM system, the location call data is very sparse, leading to around 40 tuples per mobile user per day. As a result, only few selective tuples are used for the trip analytics of a transit network. (Poonawala et al., 2016) further proposes the development of trajectory analytics for automatically calibrate the location of a call using farecard data, allowing for high spatial coverage. However, the procedure of matching a user location onto a train network is essential for this approach, hampering its use in streaming analytics, mainly because the efficiency and scalability issues of current map matching algorithms.

Other approaches for improving transit management include gathering transit feeds using Automatic Vehicle Location (AVL) or Global Positioning System (GPS) and developing architectures for batch processing analytics. The research work carried out by (El-Geneidy et al., 2011) demonstrates how AVL data can be analyzed at a microscopic level to understand causalities of delay and avoid operational problems of transit service reliability. However, the batch processing analysis is focused on the time dimension for predicting travel time, and no attention was paid to the geographical space. In contrast, (Ram et al., 2016) proposes a 3-layer web-based system for replacing AVL with GPS technology since it is a more economical alternative for traffic monitoring. The system is implemented using a Hadoop-based distributed processing to improve the running times and an open GIS for processing bus travel time and passenger boarding locations.

## 3. STREAMING ANALYTICAL WORKFLOW

We propose a streaming analytical workflow based on the main characteristics of data streams as described by (Gama and Rodrigues, 2007):

- Each tuple in a stream arrives online.
- A system has no control over the order in which a tuple arrives within a data stream or across data streams.
- Data streams are potentially unbounded in size.

The transit feeds used in this research have the above characteristics, and they are generated with a high data input rate as well. They consist of a sequence of out-of-order tuples containing attributes such as:

$$T1 = (S_1, x_1, y_1, t_1)$$

where

$S_1$ : is a set of attributes containing information about each bus of a transit network, for example: bus route identifier, bus route number, vehicle identifier, trip identifier, start time of a trip, and the end time of a trip.

$x_1, y_1, t_1$: is the geographical location of a bus of a transit network at the timestamp $t$.

Our streaming analytical workflow consists of three automated tasks including (1) data ingestion; (2) data cleaning; and (3) data contextualization. The automation of this workflow is of paramount importance to streamline and process large amount of transit feeds. Figure 1 provides an overview of these tasks which have been implemented within a Python algorithm running in a cloud computing environment. These tasks are further explained as follows.

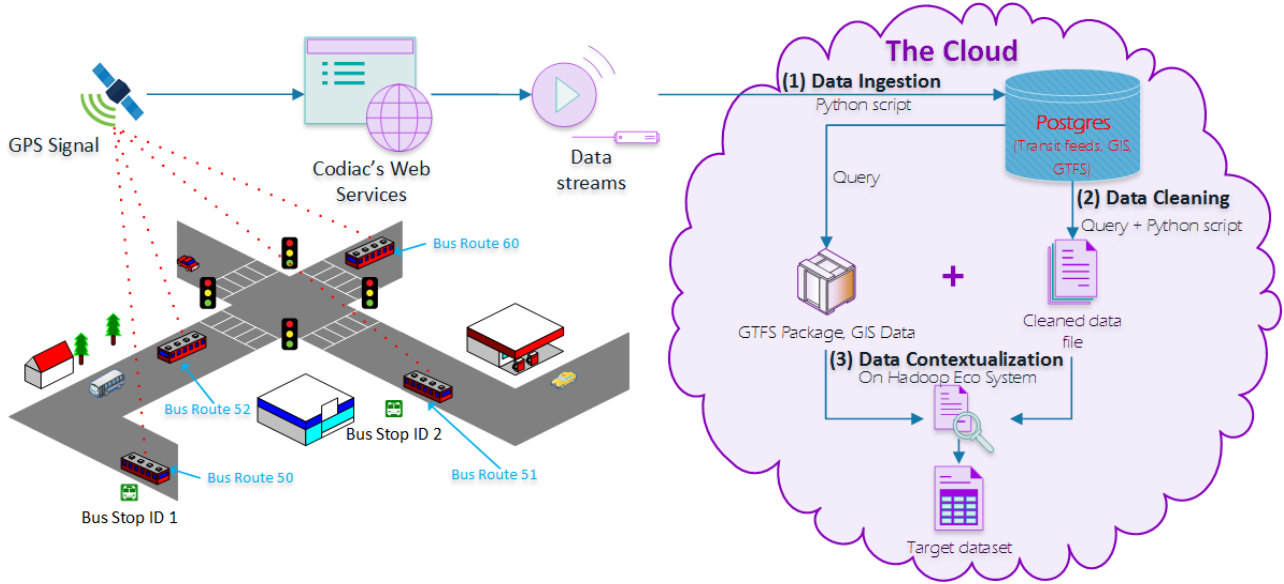

Figure 1. Overview of our streaming analytical workflow

### 3.1 Data Ingestion

Data ingestion is known as the task of pushing data streams from different sources into the cloud, which supports virtualized data processing of continuous and high volume of data streams. The Python script algorithm is implemented to allow an http POST connection for rapid collection of the data streams from the broadcasting service of a transit network in which a forever loop of event time windows of 5 seconds is applied (see Table 2). There are two advantages of using event time windows. First, it decouples the program semantics from the actual serving speed of the source and the processing performance of system. Hence historic data can be processed, which is served at maximum speed, and continuously produced data with the same program. It also prevents semantically incorrect results in case of backpressure or delays due to failure recovery. Second, event time windows compute correct results, even if the data streams arrive out-of-order of their timestamp.

All data streams are then stored in a PostgreSQL database in the cloud at a periodic interval of 5 seconds. Although NoSQL databases such as MongoDB, Cassandra, Amazon Dynamo to mention a few, are well suited for storing and indexing real-time stream spatial data, they might lead to potential schema-less problems. Specifically, the lack of a database schema may cause an application to fail due to application behaviour and expectations. The PostgreSQL database provides a central database schema in the cloud, and query language to retrieve the data streams needed for the data cleaning task. Moreover, the PostgreSQL community have added many new features and better performance for big data use cases including the ability to store unstructured data and add a column on fly in a dynamic table (Chihoub and Collet, 2016).

The PostgreSQL database is also used to store any other information required for the data contextualization task. This includes a GIS layer and the General Transit Feed Specification (GTFS) data. The GIS layer contains a 30m buffer zone along each bus route line of a transit network containing. This buffer zone consists of a square grid cells of 10m and each grid cell contains a tag with the name of the street segment that a bus route line belongs to. The GTFS (Antrim et al., 2013) is a set of files that provides a common format for public transit schedules and associated geographic information. Transit agencies can publish their transit data using GTFS and allow developers to design applications which can consume this data in an interoperable way.

The transit feeds keep being pushed to the cloud and stored in the PostgreSQL database. With in-memory processing, the PostgreSQL database is queried to retrieve the data using different time periods, ranging from hour, day, week, month and year. In this paper, we have executed a query to retrieve a year of transit data feeds, which are later used for the data cleaning step.

### 3.2 Data cleaning

Data cleaning is always necessary in order to remove errors and inconsistencies from the stored data streams. Guaranteeing data quality for continuous and high volume of data streams is a nontrivial task, and performing this task automatically is even more challenging because the streaming rate is highly dynamic. The Python script algorithm is implemented to handle five automated steps for dealing with *(1) missing tuples, (2) duplicated tuples, (3) missing attribute values, (4) redundant attributes, and (5) wrong attribute values.*

**Step 1: Missing tuples**. Every tuple is expected to arrive every 5 seconds for each bus trip. However, due to connectivity and/or sensor problems, missing tuples might occur.
E.g.:
*Bus 1- Bus Route 51 - Trip 1*
$(S_1, x_1, y_1, t_1)$; <u>missing tuple</u>; $(S_3, x_3, y_3, t_3)$; ….$(S_n, x_n, y_n, t_n)$

*Bus1 - Bus Route 51 - Trip 2*
$(S_1, x_1, y_1, t_1)$; <u>missing tuple; missing tuple</u>; ….$(S_n, x_n, y_n, t_n)$

In this case, we eliminate any bus trip that has in total 100 missing tuples and more. Any trip that has lost data for 8.33 minutes (5 seconds x 100) will be not used in the data contextualization task.

**Step 2: Duplicated Tuples.** It includes the case when the same tuple is transmitted twice. In this case, any duplicated tuple is automatically found using its timestamp and then removed.
E.g.:
*Bus 1- Bus Route 51 - Trip 1*
$(S_1, x_1, y_1, t_1)$, $(S_2, x_2, y_2, t_2)$; <u>$(S_3, x_3, y_3, t_3)$; $(S_3, x_3, y_3, t_3)$</u>, ….$(S_n, x_n, y_n, t_n)$

**Step 3: Missing attribute values.** *S* is a set of a-priori known number of attributes which is generated every 5 seconds. This set may contain several attributes such as vehicle id, bus route number, and bus route id. If the missing attribute value is not used in the further steps, the "N/A" is assigned to this attribute. Otherwise, we delete this tuple.

**Step 4: Redundant Attributes.** Although *S* has a fixed number of attributes, there are cases when a new attribute is introduced during the data ingestion task. For example, in the case of having a defined as a set of 4 attributes, it might occur that 5 attributes are retrieved instead. In this case, the extra attribute is automatically deleted.

**Step 5: Wrong attribute values**. Any attribute might also contain a wrong value due to misspelling, illegal values, and uniqueness violation. In this case, the algorithm first try to standardize the wrong information. But if the attribute cannot be standardized, the attribute value is treated as a missing attribute value case.

Once the data cleaning task is finished, the algorithm creates a data set and stores it as a HDFS file in Hadoop. This is the cleaned data sample ready to be used in the data contextualization task.

### 3.3 Data Contextualization

This is the most complex task designed for the analytical workflow. Contextualization enriches the tuples from the previous data cleaning task using higher level concepts accordingly to a particular mobility context. In our case, we are interested in the mobility context of a bus trip. The main goal is to systemize the trips of a bus route in the most effective manner for improving comprehension to further analytics. Towards this end, the data contextualization task actually consists of seven automated steps which have been implemented as illustrated in Figure 2. All the steps implemented for this task are executed using a Python script algorithm (See Table 2).

**Step 1: Stop/Move Detection.** The aim is to determine whether a bus is moving or has stopped off or has suspended its movement at a street segment or intersection. The geographical coordinates $(x, y)$ and the timestamp *t* of each tuple are used for this contextualization. First, an empirical distance value of 15m is selected to compute stops and moves. Second, the Euclidean distance between two consecutive points (i.e. tuples) is computed. If the distance between two consecutive points is larger than 15m, a new attribute containing the value "*move*" is added to the second tuple. In contrast, if the distance is less than 15m, the "*stop*" attribute value is added to the second tuple.

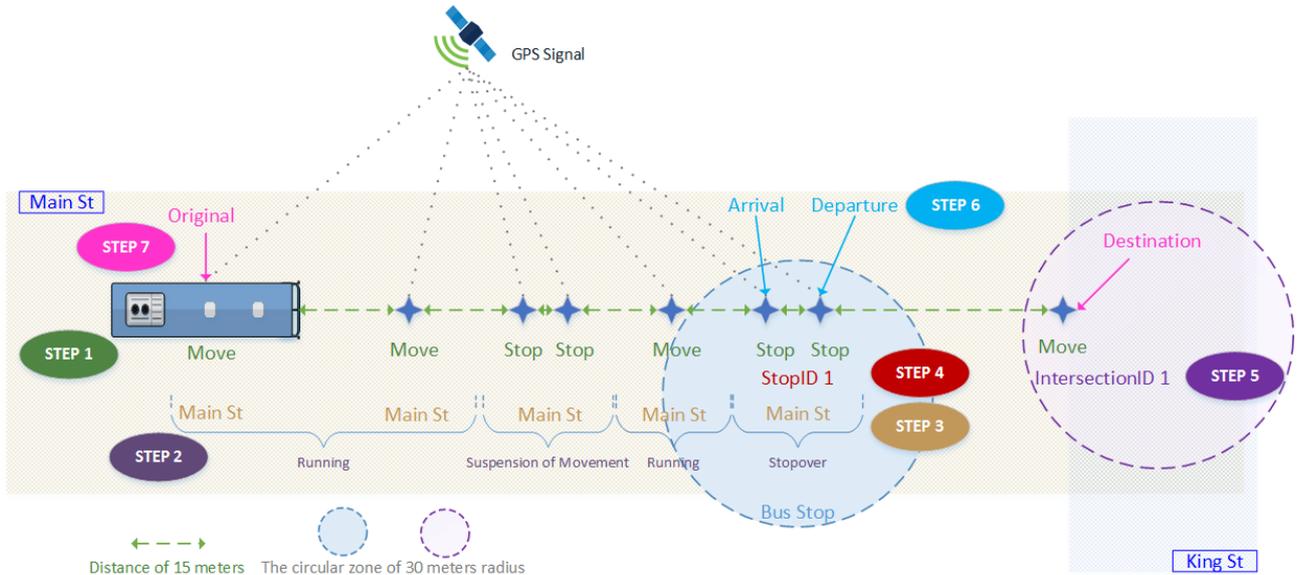

Figure 2. Overview of the automated steps designed for the data contextualization task

**Step 2: Stop/Move Classification.** The aim is to classify the moves and stops obtained from the previous step in order to improve our understanding about their mobility patterns during a trip. Any stop may occur because of a traffic jam, accident, collecting passengers at a bus station, or a traffic light at one street intersection. In this case, the classification is carried out by adding one new attribute which can contain one of the following values:

1. Running: when a bus is moving on a street;
2. Passing: when a bus passes a bus station because there are no passengers to drop off or get on.
3. Suspension of movement: It may occur due to an intersection, stop sign, accident, or traffic jam.
4. Stopover: when a bus stops at a bus station for dropping off or picking up passengers.

Therefore, a query is executed to retrieve the geographical location of all the bus stations of a bus route from the PostgreSQL database. This information is available from the GTFS data previously stored in PostgreSQL database. After, the algorithm creates a circular zone with a radius of 30m for each bus station. The stops which are located inside the buffer are classified as *"stopovers"*, otherwise, they are classified as *"suspension of movement"*. Moreover, the moves which are located inside the buffer are classified as *"passing"*, otherwise they are classified as *"running"* on a street.

**Step 3: Street Name Annotation.** The aim is to annotate the moves and stops accordingly to their spatial semantics, in particular, using the streets nomenclature of a transit network. For this step, a query is run to retrieve the names of the street where a move or stop is located at. Therefore, the GIS layer already stored in the PostgreSQL database is used for the contextualization.

This is a non-trivial step because the geographical coordinates of the stops and moves are obtained from GPS signals which can range from 10m to 100m accuracy in urban areas (Salarian et al., 2015). The grid-based buffer zone plays an important role in indexing which street segment any cell belongs to, and after localize the moves and stops within a cell, and consequently, identify the street name.

**Step 4: Bus Station Identification.** The aim is to tag a bus station id to each tuple containing the attribute values equal to stopover and passing. This is an important step to provide a link with the bus station id information available from the GTFS data. This is achieved by creating a circular zone of a 30m radius around each bus station of a transit network, and matching it with the stop (i.e. stopover and passing) location of a moving bus. It is important to point out that the algorithm also needs to verify the direction of a moving bus (e.g. eastbound and westbound) in order to identify the bus station that a stopover/passing is actually located. We select a tuple located at the middle of a bus route for using it as a reference point for identifying the direction of a moving bus. Each stop can be then annotated using "outbound" and "return" values.

**Step 5**: **Street Intersection Identification**. The aim is to tag an intersection id to each tuple. The algorithm creates a circular zone with a radius of 30m for each street intersection. The tuples containing stops and moves that are located inside the circular zone are tagged with the intersection id. Otherwise, the NULL value is used.

**Step 6: Arrival/Departure Times Identification.** The aim of this step is to determine the actual arrival time and departure of a moving bus for dropping off or picking up passengers. In this case, the algorithm verifies for the timestamp of the first stopover within the circular zone of 30m radius around each bus station, and considers it as the actual arrival time. Similarly, the timestamp of the last stopover within the circular zone is considered the departure time. This step can be improved if automatic passenger counters (APCs) are used in a transit network because they provide information about passenger activity on bus trip time.

**Step 7: Origin/Destination Trip Identification**. The aim is to tag each first tuple of a bus trip as origin, and each last tuple of a bus trip as destination. The other tuples are then sequentially indexed.

At the end of the contextualization task, a contextualized data sample is generated and stored as a HDFS file in Hadoop. This sample contains seven new attributes added to the original set of tuples obtained from the data cleaning task. These attributes actually represent the mobility context that characterises the situation, environment, and surroundings that are related to a bus trip of a transit network. Moreover, sampling in our streaming analytical workflow is used to reduce the input rate of the data streams for preserving high processing capacity to further analyse the tuples using machine learning and pattern discovery techniques. But this further analysis is out of the scope of this paper.

## 4. STREAM PROCESSING ARCHITECTURE

Stream processing is required for supporting the continuous computation of data flowing through our analytical tasks. When any tuple is retrieved by the Hadoop system, its attributes are processed as soon as they arrive. The only constraint is that the output rate should be at least similar to the data input rate, mainly due to have enough memory to store the cleaned and target data samples from the analytical workflow.

Hadoop is used for processing the data streams. It is a Java-based open-source software framework that supports distributed storage and distributed processing of very large datasets across computer clusters of commodity servers using the MapReduce programing model (Dean and Ghemawat, 2010). Hadoop was selected because it is designed to run applications on systems that could scale up from a single machine to thousands of computers, with very high level of fault tolerance. The distributed file system (HDFS) facilitates rapid data transfer rates among machines and allows the system to keep working uninterrupted in case of a server failure. It divides HDFS data into large blocks that can be handle on many servers in the cluster. To process the data, Hadoop framework transfers packaged code for machines to process in parallel, based on the data each machine needs to process.

In our system, a Hadoop cluster includes a master node and a server node that were deployed using the Compute Canada West Cloud following the specifications listed in Table 1. Figure 3 shows an overview of the architecture used to process one year of data streams.

| Hadoop cluster | |
|---|---|
| Master | Hostname: first-hung.westcloud <br> OS: CentOS 7.0 (x86_64) <br> CPU: Intel(R) Xeon(R) CPU E5-2650 v2 @ 2.60GHz <br> Number of CPU core: 8 <br> RAM: 29.3 GB <br> Disk: 859 GB <br> IPv4 Address: 192.168.14.60 |
| Slave | Hostname: third-hung.westcloud <br> OS: CentOS 7.0 (x86_64) <br> CPU: Intel(R) Xeon(R) CPU E5-2650 v2 @ 2.60GHz <br> Number of CPU core: 8 <br> RAM: 29.3 GB <br> Disk: 859 GB <br> IPv4 Address: 192.168.14.67 |

Table 1. The Hadoop Specifications

The MapReduce programing model in our workflow performs two functions. First, the Map phase divides the original tuple into key-value pairs then shuffle into many small subsets with the same key. Second, the map phase maps this input data to a set of intermediate key/value pairs as follows.

$$Map\ (Dataset) \rightarrow [K;\ list(V_1)]$$

The *Route_ID*, *Trip_ID*, and *Date (year:month:day)* attributes of each tuple are used to establish the key for this function. After executing the map phase, the result is a key/value pairs in which the value is the list of sorted many tuples T that have the same key as follow.

$$Map\ (T_1,…,T_n) \rightarrow [(Route\_ID, TripID, Date); list(sorted\_subset\ (T))]$$

The Reduce phase takes these subsets and applies the same processing algorithm in a parallel manner to produce a single result set. Reduce phase reduces a set of intermediate values which share a key *K* to a smaller set of values *list ($V_2$)* as follow.

$$Reduce\ ([K;\ list(V_1)]) \rightarrow list(V_2)$$

The input of the Reduce function is used to partition the tuples into groups of trips belonging to the same bus route at different days. The contextualization task is then run for each partition containing all the trips belonging to the same bus route (i.e. Reduce Task 1). At the end of the Reduce phase, all output of the Reduce function is grouped into a list of processed tuples of the form

$F_1(S_1, a_{18}, a_{19}, a_{20}, a_{21}, a_{22}, a_{23}, a_{24})$;
$F_2(S_2, a_{18}, a_{19}, a_{20}, a_{21}, a_{22}, a_{23}, a_{24})$;
…..
$F_n(S_n, a_{18}, a_{19}, a_{20}, a_{21}, a_{22}, a_{23}, a_{24})$

where $a_{18}, a_{19}, a_{20}, a_{21}, a_{22}, a_{23}, a_{24}$ are the new attributes. The logical view of this phase is shown in Figure 3.

$$Reduce\ ([(Route\_ID, TripID, Date); list(sorted\_subset\ (T))]) \rightarrow List(F(a_{18}, a_{19}, a_{20}, a_{21}, a_{22}, a_{23}, a_{24}))$$

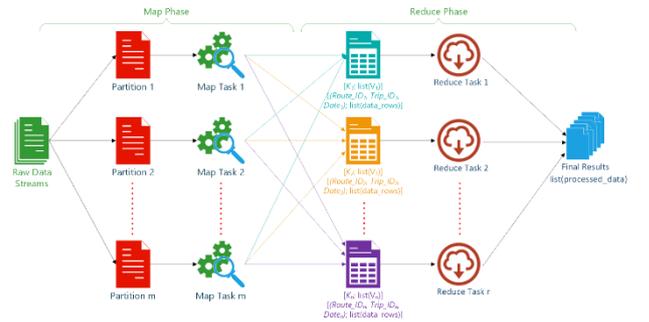

Figure 3. MapReduce Programing Model

## 5. IMPLEMENTATION

The CODIAC transit network for Greater Moncton area was used for the implementation of our streaming analytical workflow. It has 30 bus routes operating from Monday to Saturday, some of which provide evening and Sunday services. Every bus in the transit network has installed a GPS receiver for collecting its location every 5 seconds. The set of attributes (S) consists of:

1. *vlr_id: The ID of the data point in the vehicle location reports table*
2. *route_id_vlr: The route ID in the vehicle location reports table*
3. *route_name: The route name*
4. *route_id_rta: The route ID in the route transit authority table*
5. *route_nickname: The abbreviate of the route*
6. *trip_id_br: The trip ID in the bid route table*
7. *transit_authority_service_time_id: Transit authority service time ID*
8. *trip_id_tta: Transit authority trip ID*
9. *trip_start: Start time of the trip*
10. *trip_finish: Finish time of the trip*
11. *vehicle_id_vab: Vehicle ID*
12. *vehicle_id_vlr: Vehilce ID in the vehicle location reports table*
13. *vehicle_id_vlr_ta: The descriptive name of the bus*
14. *bdescription: Bus description*

plus

15. *lat: Latitude*
16. *lng: Longitude*
17. *timestamp: Timestamp of the data point*

Moreover, the GTFS package that contains the scheduled stop times of every bus, all bus stations locations, routes, trips and other information of the Greater Moncton area is utilized in this implementation. In total, there are about 800 trips for the 30 bus routes that are operated each day during the period of one year. The transit network has 642 bus stations. In addition, the geographical information of Moncton Transit network is also available including the GIS shape file of the bus routes network, of the bus station locations, of the road network of Moncton City. Figure 4 illustrates the spatial distribution of the bus routes.

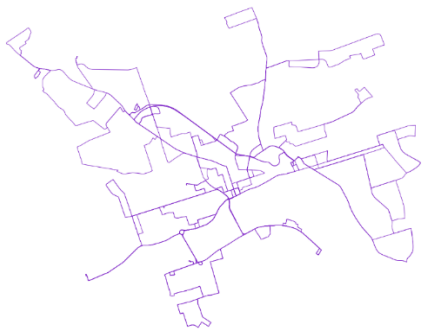

Figure 4. Spatial distribution of the CODIAC Transit Network

The pseudo code of the Python algorithm implemented to execute the tasks can be found in Table 2 below.

```
Python Algorithm
-- Data Ingestion
data_ingestion(incoming_streams)
{
  while (1) //loop forever
  {
    establish_connection(incoming_streams)
    tuples = read(incoming_streams)
    store_to_PostgreSQL(tuples)
    delay(5s) // ingest data streams every 5 seconds
  }
}

-- Data Cleaning
data_cleaning()
{
  raw_tuples = query_from_PostgreSQL(data)
  for each Trip in raw_tuples:
    check_missing_tuple(Trip)
    data_1 = clean_missing_tuple(Trip)
    check_missing_attribute(data_1)
    data_2 = fix_missing_attribute(data_1)
    check_wrong_attribute(data_2)
    data_3 = fix_wrong_attribute(data_2)
    check_redundant_attribute(data_3)
    data_4 = eliminate_redundant_attribute(data_3)
    check_duplicated_tuple(data_4)
    cleaned_data = eliminate_ duplicated_tuple(data_4)
  return cleaned_data
}

-- Data Contextualization which is performed on Hadoop Eco System using MapReduce programming model.
data_contextualization()
{
  map_phase(cleaned_data)
  {
    for each tuple in cleaned_data
      key = <Route,Trip,Date>
      value = each tuple
    shuffle(key, value)  //sort all tuples with the same key to
                        //small subsets
    return <key,value> //many subsets (Trips)
  }

  reduce_phase(<key,value>)
  {
    for each Trip in <key,value>
      apply_seven_contextualization_steps(Trip)
      final_result = join_the_result_together(Trip)
    return final_result
  }
}
```

Table 2. The pseudo code of the workflow tasks

**Data Ingestion**

The transit data streams have been continuously being pushed to our PostgreSQL database on the west cloud of Compute Canada since 01/06/2016. We retrieved a year of streaming data from 01/06/2016 to 25/05/2017 to test the proposed analytical workflow. At the time, there were 65,097,658 tuples stored in the PostgreSQL database that were used for the data cleaning task.

**Data Cleaning**

Errors and inconsistences information needed to be corrected and filtered out. In the case of *missing tuples,* 480,000 tuples were deleted accounting for 0.75% of total of 65,097,658 tuples, and because 6,000 bus trips had more than 100 missing tuples, they have been removed as well.

Furthermore, around 6,000 tuples were standardized due to the cases of *missing attribute values, redundant attributes, and wrong attribute values*. Finally, 38,167,787 tuples were detected to be *duplicated tuples,* and consequently, they have been deleted as well. At the end, the cleaning data file consisted of 26,443,871 tuples which were used for the data contextualization task.

**Data Contextualization**

The cleaned tuples were then contextualized following the seven steps described in Section 3.3. The whole streaming processing of the tuples handled by these tasks have been done in our Hadoop system described in Section 4. The bus route 51-12 on the date 15-06-2016 was randomly selected for illustrating the outcomes of the steps.

In the first step of contextualization task, moves and stops are computed. Figure 5 shows one bus trip of the bus route 51 after this computation. This trip had 518 tuples, having 230 moves and 288 stops.

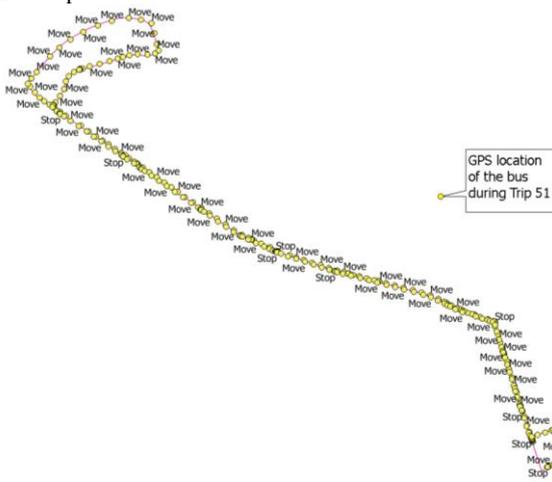

Figure 5. Step 1 – Stop/Move Detection

In step 2, the moves and stops belonging to this bus trip were classified as running, passing, suspension of movement and stop over. Figure 6 illustrates the results found where 200 moves were classified as running and 30 moves as passing while 62 stops were classified stopover and 226 stops as suspension of movement.

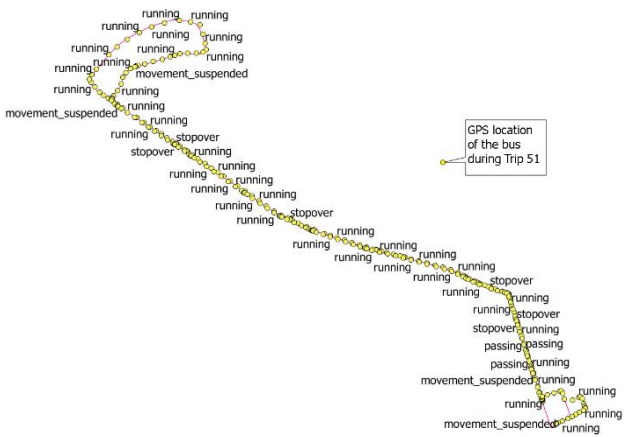

Figure 6. Step 2 – Stop/Move Classification

The step 3 of the data contextualization task was implemented by using the GIS layer containing a 30m buffer zone along each bus route line of a transit network (See Figure 4 for the whole transit network). Figure 7a shows the bus route 51 and Figure 7b shows an example of the grid-based buffer zone used for retrieving the street names. In this case, it is possible to see that the moving bus has not followed the assigned bus route. This might have occurred due to an accident, road construction or any other event that required the driver to drive on different street segments.

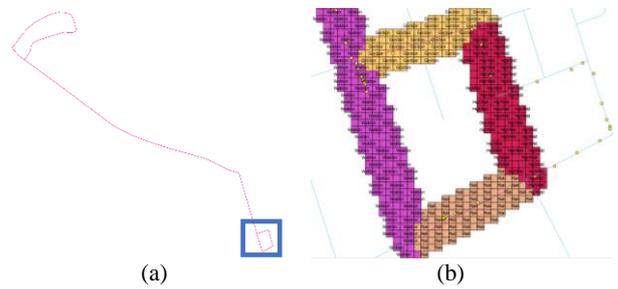

(a)          (b)

Figure 7. Example of the GIS layer containing the 30m buffer zone along bus route 51

Figure 8 describes the results after the tuples are enriched with the street segment's name. In the case a moving bus does not follow the designated street segment, the algorithm generates the "wrong street segment" value.

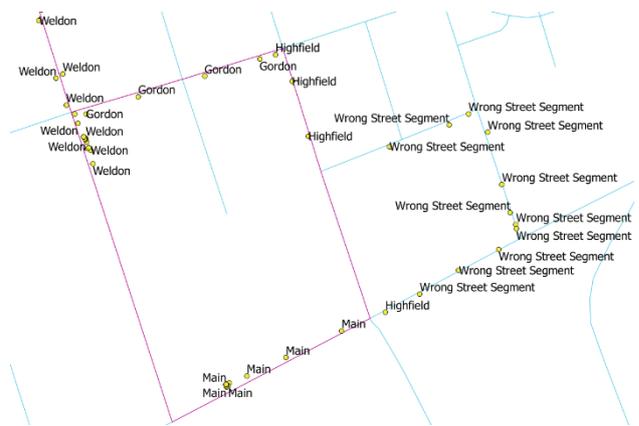

Figure 8. Step 3 – Street Name Annotation

Next, the mobility context is inferred by identifying the physical stop of a moving bus. Using the GTFS data stored in the PostgreSQL database, the location of a bus station is compared with an actual stop of a moving bus (Figure 9).

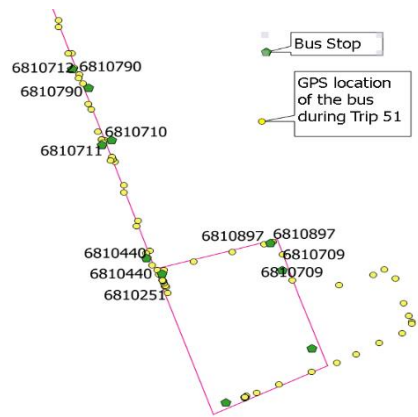

Figure 9. Step 4 – Bus Station Identification

Figure 10 shows the stops annotated as outbound and returns of the trip for the bus route 51.

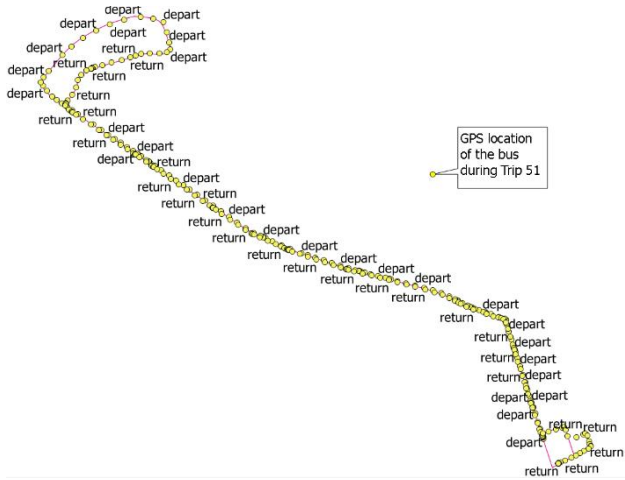

Figure 10. Outbound and return of the stops and moves near a physical bus station

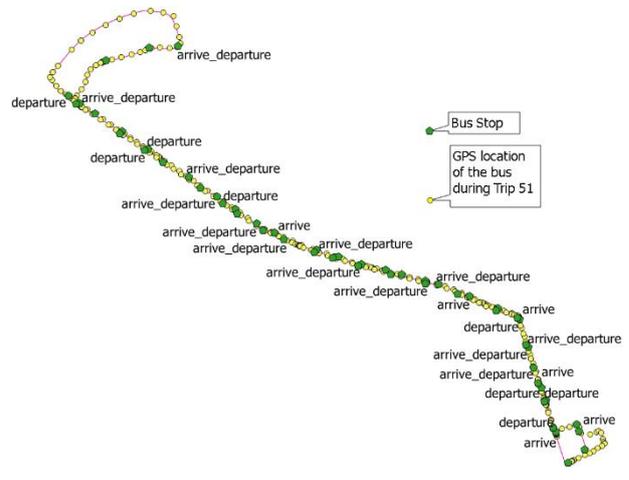

Figure 12. Step 6 – Arrival/Departure Times Identification

In step 5, the stops and moves close to an intersection are identified (see Figure 11).

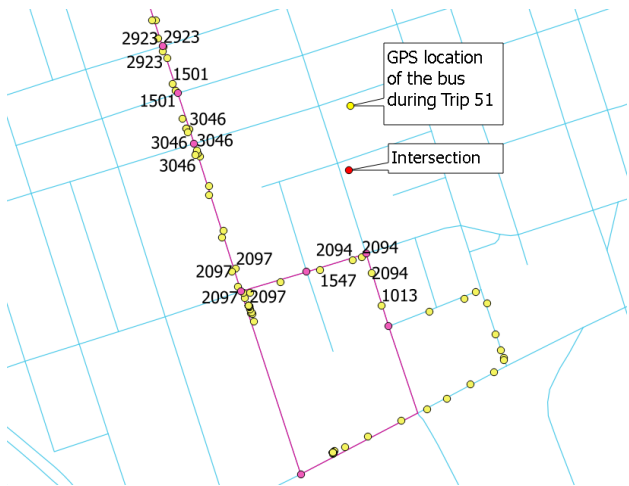

Figure 11. Step 5 – Street Intersection Identification

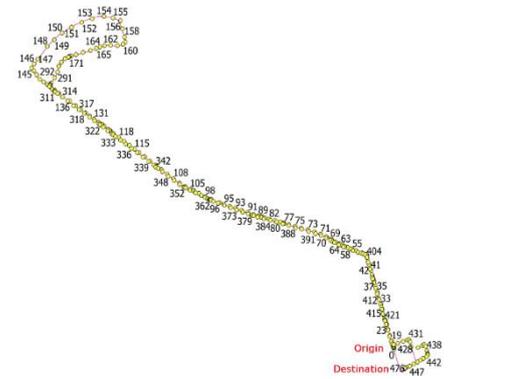

Figure 13. Step 7 – Origin/Destination Trip Identification

In Step 6, the arrival and departure times of a bus station over are computed as illustrated in Figure 12, and finally the origin and destination stops are identified as illustrated in Figure 13. The results of this step show that the origin and destination of this particular trip were note located at the same bus station.

Aiming to evaluate the automatic batch processing of the data contextualization task using MapReduce, two datasets were extracted from the cleaned tuples to run in Hadoop. The first dataset A contains the 12.75 million cleaned tuples from 01/06/2016 to 15/12/2016. The second data set B contains 13.69 million cleaned tuples from 16/12/2016 to 25/05/2017. Figure 14 shows the processing time for all phases including map phase, shuffle phase, and reduce phase. Notably, the Reduce processing time is much longer than Map processing time because the Reduce phase runs all the data contextualization steps while the Map phase mainly sorts tuples into separate cluster of the same bus route.

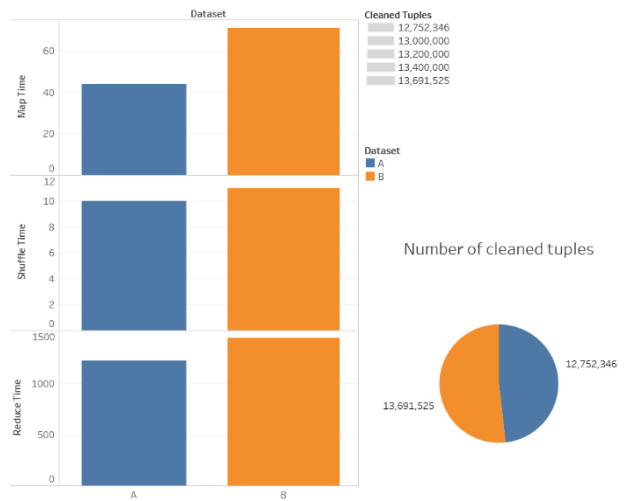

Figure 14. MapReduce processing time

## 6. CONCLUSION

Developing a streaming analytical workflow requires understanding the structure of data streams together with the software architectures needed for processing them. This is not a trivial activity, mainly because any streaming analytical workflow will consist of many tasks which will require several automated steps. In the past, the analysis of transit feeds has been focussed on the temporal semantics of the transit trips such as run time, headway deviation and number of scheduled stops. Our streaming analytical workflow is paving the way to show that is is possible to explore the spatial semantics of a transit network as well. It will require high performance computing power to support the contextualization tasks of these workflows, but the transit feeds will be contextualized with information about the mobility context of *where and when* the transit trips are taking place in the network. Analytics performed over contextualized transit feeds could potentially revolutionize transit network services that will be able to adapt at near real time to current or expected mobility contexts, implementing real-time operation controls and recommender systems.

However, the outcomes from the data cleaning task indicate that it is not worth it to send all the data streams to the cloud since most of them will not be used in the contextualization task. Almost half of the tuples used in our implementation were deleted during the data cleaning task. Other computing architectures such as mobile fog computing might be more appropriate for performing the data cleaning task at the edge of the network, rather than at the cloud. Mobile fog computing is defined as "a scenario where a huge number of heterogeneous (wireless and sometimes autonomous) ubiquitous and decentralized devices communicate and potentially cooperate among them and with the network to perform storage and processing tasks without the intervention of third-parties (Vaquero and Rodero-Merino, 2014). Data cleaning tasks can be designed for running in a sandboxed environment at a fog node. Future research work includes to implement the data cleaning task at a mobile fog node which would be installed inside a vehicle of a transit network.

Finally, it is important to point out that the 30m circular zones might not be a universal radius to be adopted by any transit network. More research work is needed to identify the optimal radius value for the circular zones used for bus stations and intersections.

## ACKNOWLEDGEMENTS

This research was supported by the NSERC/Cisco Industrial Research Chair in Real-Time Mobility Analytics. The authors are grateful to CODIAC Transit for providing the data streams used in this study, and Compute Canada for hosting two virtual machines that were used for running the Hadoop cloud-based architecture. The authors would also like to thank Ryan Brideau, Iyke Maduako and Emerson Cavalleri from the People in Motion Lab for their constructive comments on the development of the streaming analytics workflow.